\documentclass[fleqn,usenatbib]{mnras}

\usepackage[T1]{fontenc}
\usepackage{ae,aecompl}
\usepackage{graphicx}	
\usepackage{amsmath}	
\usepackage{amssymb}	

\title[Turbulence driving mode in molecular clouds]{On the effective turbulence driving mode of molecular clouds formed in disc galaxies}

\author[K.~Jin et al.]{
Keitaro Jin$^{1}$,
Diane M.~Salim$^{2,3}$,
Christoph Federrath$^{2}$\thanks{E-mail: \href{mailto:christoph.federrath@anu.edu.au}{christoph.federrath@anu.edu.au}},
Elizabeth J.~Tasker$^{1,3}$,
\newauthor{ 
Asao Habe$^{1}$,
Jouni T.~Kainulainen$^{4}$}
\vspace{0.2cm}\\
$^{1}$Department of Physics, Faculty of Science, Hokkaido University, Kita 10 Nishi 8 Kita-ku, Sapporo 060-0810, Japan\\
$^{2}$Research School of Astronomy and Astrophysics, Australian National University, Canberra, ACT 2611, Australia\\
$^{3}$Institute of Space and Astronomical Science, Japan Aerospace Exploration Agency, Yoshinodai 3-1-1, Sagamihara, Kanagawa, Japan \\
$^{4}$Max Planck Institute for Astronomy, K\"onigstuhl 17, 69117 Heidelberg, Germany\\
\vspace{-0.3cm}\\For enquiries please email the corresponding author: \href{mailto:christoph.federrath@anu.edu.au}{christoph.federrath@anu.edu.au}
}

\begin{document}

\maketitle

\begin{abstract}
We determine the physical properties and turbulence driving mode of molecular clouds formed in numerical simulations of a Milky Way-type disc galaxy with parsec-scale resolution. The clouds form through gravitational fragmentation of the gas, leading to average values for mass, radii and velocity dispersion in good agreement with observations of Milky Way clouds. The driving parameter ($b$) for the turbulence within each cloud is characterised by the ratio of the density contrast ($\sigma_{\rho/\rho_0}$) to the average Mach number ($\mathcal{M}$) within the cloud, $b=\sigma_{\rho/\rho_0}/\mathcal{M}$. As shown in previous works, $b\sim1/3$ indicates solenoidal (divergence-free) driving and $b\sim1$ indicates compressive (curl-free) driving. We find that the average $b$ value of all the clouds formed in the simulations has a lower limit of $b > 0.2$. Importantly, we find that $b$ has a broad distribution, covering values from purely solenoidal to purely compressive driving. Tracking the evolution of individual clouds reveals that the $b$ value for each cloud does not vary significantly over their lifetime. Finally, we perform a resolution study with minimum cell sizes of $8$, $4$, $2$ and $1\,\mathrm{pc}$ and find that the average $b$ value increases with increasing resolution. Therefore, we conclude that our measured $b$ values are strictly lower limits and that a resolution better than $1\,\mathrm{pc}$ is required for convergence. However, regardless of the resolution, we find that $b$ varies by factors of a few in all cases, which means that the effective driving mode alters significantly from cloud to cloud.

\end{abstract}


\section{Introduction}

Star formation occurs in the coldest phase of the interstellar medium (ISM), where the gas forms extended structures known as giant molecular clouds (GMCs). However, the internal mechanisms within the clouds that produce a galaxy's star formation rate remain poorly understood. The clouds themselves cannot be freely collapsing to form stars, since the rate of gas conversion over a cloud's free-fall time would lead to a star formation rate more than a hundred times higher than observed in the Milky Way \citep{ZuckermanEvans1974,WilliamsMcKee1997,KrumholzTan2007}. Two main options exist for avoiding this conundrum. The first is that the clouds are destroyed by its first generation of stars, redistributing the majority of its gas back into the warm ISM \citep{Murray2011}. However, previous simulations have failed to successfully disperse the clouds with stellar feedback \citep{HopkinsQuataertMurray2012,TaskerWadsleyPudritz2015,HowardPudritzHarris2016}. While it is possible that the feedback methods employed within the simulations are ineffective for numerical reasons, there is also strong observational evidence to suggest that clouds are not in free-fall. Rather, internal turbulent motions and magnetic fields seem to control the dynamics of the clouds and prevent the rapid conversion of gas into stars \citep{FederrathKlessen2012,PadoanEtAl2014}. 

Support for clouds being turbulent structures stretches back to the early 1980s with the seminal paper by \citet{Larson1981}. Since then, it has become widely accepted that GMCs are supersonically turbulent and this must play a crucial role in star formation \citep{Padoan1995,ElmegreenScalo2004,MacLowKlessen2004,McKeeOstriker2007,HennebelleFalgarone2012,PadoanEtAl2014}. This then leads to the question of the origin of this turbulence. Without a fresh injection of energy, turbulent modes inside a cloud will decay over a crossing time, $t_{\rm cross} \sim L/\sigma \sim 20\,{\rm pc}/5\,{\rm km\,s^{-1}} \sim 4\,{\rm Myr} \sim t_\mathrm{ff}$ and the clouds would quickly collapse and form stars at a high rate \citep{StoneOstrikerGammie1998,MacLowEtAl1998,MacLow1999}. A force is therefore needed to drive the turbulence in the clouds, either from the inside through stellar feedback, or from the outside via shear  or interactions with neighbouring clouds \citep{FederrathEtAl2017iaus}. Evidence for the latter comes from previous simulations of cloud formation. Even without the inclusion of star formation, global simulations forming clouds through self-gravitational collapse find good agreement with observations for the typical properties such as mass, radius and velocity dispersion \citep[]{TaskerTan2009,BenincasaEtAl2013,FujimotoEtAl2014}. Including turbulence in analytical theories of star formation also produces a good match with observed star formation rates and predictions of the initial mass function of stars \citep[]{PadoanNordlund2002,HennebelleChabrier2011,HennebelleChabrier2013} and star formation rates and efficiencies \citep[]{KrumholzMcKee2005,PadoanNordlund2011,HennebelleChabrier2011,FederrathKlessen2012,FederrathKlessen2013,Federrath2013sflaw,PadoanEtAl2014,SalimFederrathKewley2015}. This suggests that GMC evolution may be largely governed by gravity and turbulence driven by interactions within their galactic environment. Indeed, \citet{TaskerTan2009} and \citet{FujimotoEtAl2014} find that cloud collisions are common, occurring multiple times per orbital period. Observations also support this view, with suggestions that the turbulence itself is dominated by larger-scale modes and the existence of clouds such as the Maddalena Cloud and Pipe Nebula which have a low star formation rate \citep{OssenkopfMacLow2002,BruntHeyerMacLow2009,HughesEtAl2010}. To understand how GMCs convert their gas into stars, we must first understand the properties of the turbulence.

Information about the turbulence can be found in the gas probability density function (PDF). In an isothermal system, the gas density PDF has a log-normal form \citep[]{Vazquez1994}, with the width of the PDF increasing with Mach number; $\sigma_{\rho/\rho_{0}}=b\mathcal{M}$, where $\sigma_{\rho/\rho_0}$ is the standard deviation of the gas density, $\rho$, and $\rho_0$ is the mean density, $\mathcal{M}$ is the RMS Mach number of the turbulence and $b$ is a proportionality factor related to whether the turbulence is driven by a solenoidal (divergence-free) driver or by a compressive (curl-free) driver \citep[]{PadoanNordlundJones1997,FederrathKlessenSchmidt2008,FederrathDuvalKlessenSchmidtMacLow2010,PriceFederrathBrunt2011,PadoanNordlund2011,MolinaEtAl2012,KonstandinEtAl2012ApJ,Federrath2013,NolanFederrathSutherland2015,FederrathBanerjee2015}. Solenoidal and compressive driving of the turbulence gives rise to extremely different density distributions and the value of $b$ is determined by the relative mix of the two. In the case of fully solenoidal driving, the proportionality term becomes $b=1/3$, while fully compressive driving yields $b=1$  \citep[]{FederrathKlessenSchmidt2008,FederrathDuvalKlessenSchmidtMacLow2010}.

In a $250\,\mathrm{pc}$ sized simulation of supernova-driven turbulence performed by \citet{PadoanEtAl2016}, \citet[]{PanEtAl2016} found a ratio of compressible to solenoidal velocity modes of about $0.3$, which \citet[]{PanEtAl2016} suggested may relate to an average $b\sim0.5$. This suggests a significant fraction of compressible modes in the turbulence driving. Observationally, \citet{Brunt2010} estimated $b \sim 0.5$ from ${}^{13}$CO line observations of the Taurus molecular cloud, using the projected 2D column density. \citet{KainulainenTan2013} found a slightly lower (more solenoidal) value of $b \sim 0.20^{+0.37}_{-0.22}$, for the Milky Way's infrared dark (non-star forming) clouds, however, they did not take into account the magnetic pressure contribution to the standard deviation--Mach number relation. This can result in an underestimate of $b$ \citep{PadoanNordlund2011,MolinaEtAl2012}. \citet{GinsburgFederrathDarling2013} also observed a non-star forming cloud and concluded the turbulence was compressively driven with a lower limit of $b > 0.4$. Most recently, \citet{FederrathEtAl2016} determined the $b=0.22\pm0.12$ parameter in the central molecular zone (CMZ) cloud G0.253+0.016 `Brick', indicating primarily solenoidal driving due to strong shearing motions in the CMZ. This is in contrast to clouds in the galactic disc for which $b\sim0.5$, indicating strong contributions from compressive modes in the turbulence driving \citep[see review of the currently available disc clouds for which estimates of $b$ are available in][]{FederrathEtAl2016}.

In this paper, we explore the turbulence properties of clouds formed in a high-resolution (minimum computational cell size of $\Delta x \simeq 1$\,pc) galactic-scale simulation. Our main motivation is to explore how turbulence that is driven on the galactic scale through disc shear and cloud interactions travels down to the cloud scale. By measuring the value of $b$ in the resultant cloud population, we explore the relative strengths of the solenoidal and compressive modes and how these vary between the clouds and over their lifetime. No initial $b$ parameter is imposed on the simulation and the turbulence is allowed to develop self-consistently through hydrodynamical and gravitational interactions in the disc. 

Section~\ref{sec:methods} explains the galaxy simulation and the methods for cloud identification and tracking, Section~\ref{sec:results} presents our results and in Section~\ref{sec:conclusions}, we summarise our conclusions.

\section{Numerical Methods}
\label{sec:methods}

\subsection{Simulation properties and initial conditions}
\label{sec:code}

Our simulation was performed using {\it Enzo}; a three-dimensional adaptive mesh refinement (AMR) hydrodynamics code \citep{BryanNorman1997,Bryan1999,BryanEtAl2014}. The galaxy is simulated in a three-dimensional box of side $32\,\mathrm{kpc}$ with a root grid of $128^3$ cells and eight levels of refinement, giving a limiting resolution of $\Delta x_{\rm min} \lesssim 1.0\,\mathrm{pc}$. Cells were refined whenever the Jeans length dropped to less than four times the cell size \citep{TrueloveEtAl1997} or whenever the density contrast between neighbouring cells exceeded a factor of 10. To ensure the galaxy is stable in the absence of cooling, the whole disc region is kept at a minimum of two levels of refinement in addition to the root grid ($\Delta x_{\rm min} \simeq 62$\,pc) throughout the simulation, over a height of 1\,kpc above and below the disc. Within the radii of our main region of analysis (see below) and up to 100\,pc from the mid-plane, all gas is kept refined to at least level three ($\Delta x_{\rm min} \simeq 31$\,pc). The adaptive mesh then acts to refine 23\% of the gas to level four ($\Delta x_{\rm min} \simeq 15$\,pc), 10\% of the gas to level five ($\Delta x_{\rm min} \simeq 8$\,pc), 4\% to level six ($\Delta x_{\rm min} \simeq 4$\,pc), 1.3\% of the gas to level seven ($\Delta x_{\rm min} \simeq 2$\,pc) and 0.5\% of the gas to level eight ($\Delta x_{\rm min} \simeq 1$\,pc). However, within the clouds themselves, 89\% of the gas is covered by level eight.

{\it Enzo} evolves the gas using a three-dimensional version of the Zeus hydro-scheme, which uses an artificial viscosity term to handle shocks \citep[]{StoneNorman1992a}. The variable associated with this is the quadratic artificial viscosity term and was set to its default value of $2.0$.  

The gas cools radiatively to $10\,\mathrm{K}$ using a one-dimensional cooling curve created from the CLOUDY package's cooling table for metals and {\it Enzo}'s non-equilibrium cooling rates for atomic species of hydrogen and helium \citep[]{FerlandEtAl1998, Abel1997}. This minimum value is close to that of the GMCs. At this temperature, the sound speed of the gas is $0.33\,\mathrm{km}\,\mathrm{s}^{-1}$. The weakness of this model is that the gas is assumed to be optically thin and non-molecular; both untrue within the GMCs. For identifying the GMCs themselves, the lack of molecular gas is not necessarily an issue, since it is thought to be a good tracer the overall gas distribution \citep{GloverFederrathMacLowKlessen2010}. The problem with the cloud internal dynamics is fundamentally a problem with resolution which we discuss more thoroughly at the end of this paper. 

To prevent unresolved gas from collapsing, a pressure floor is imposed when the Jeans length becomes less than four times the cell size \citep{TrueloveEtAl1997}. When this operates, an artificial pressure is calculated which results in a change to the velocity flow into that cell, suppressing collapse. The true pressure and internal energy is unchanged, so this adjustment is not advected through the simulation but calculated locally when required. The disc also includes a radially dependent photoelectric heating described in the detail in \citet{Tasker2011}, but does not include any star formation, stellar feedback or magnetic fields. The simulation therefore concentrates on the early stages of star formation and the environment of the young gas cloud evolution. 

To minimise artefacts from the orientation of the Cartesian mesh, the simulation sets a co-rotation point at a radius of $6\,\mathrm{kpc}$ \citep{BenincasaEtAl2013}. At this radius, gas does not move with respect to the grid, removing potential numerical artefacts during the circular gas flow through the Cartesian system. To maximise our efforts on this region, the highest refinement level is confined to an annulus of thickness $2\,\mathrm{kpc}$ over the co-rotation point, from $r = 5$ to $7\,\mathrm{kpc}$. Our clouds are therefore analysed from inside this region.  This refinement strategy allows us to achieve the best possible resolution for the clouds and at the same allows us to maintain realistic boundary conditions by modelling an entire galactic disc.

The galaxy itself is an isolated Milky Way-type disc, with a flat rotation curve and circular velocity of $200$\,km\,s$^{-1}$. As the disc cools, it gravitationally fragments as the Toomre $Q$ parameter drops below the stability threshold. Toomre $Q$ is defined as $Q = \kappa \sigma_\mathrm{g}/(\pi G \Sigma_\mathrm{g})$, where $\kappa$ is the epicyclic frequency, $\sigma_\mathrm{g}$ is defined as $\sigma_\mathrm{g} \equiv \sqrt{\sigma_\mathrm{nt}^2 + c_\mathrm{s}^{2}}$ where $\sigma_\mathrm{nt}$ is the 1D velocity dispersion of the gas motion in the plane of the disc and $c_\mathrm{s}$ is the sound speed (our discs are initialized with $\sigma_\mathrm{nt} = 0$), and $\Sigma_\mathrm{g}$ is the gas surface density. A value $Q > 1$ corresponds to a gravitationally stable disc, while $Q < 1$ suggests an unstable state. In our main disc region between $r \sim 2$--$10\,\mathrm{kpc}$, we set the $Q$ initially to $Q = 1$. As the disc cools, the gas fragments into the clouds, which can then grow and evolve through mergers and interactions. The initial fragmentation and evolution of the gas disc is described in \citet{TaskerTan2009,Tasker2011,TaskerWadsleyPudritz2015,BenincasaEtAl2013}. By modelling the global disc, the effect of galactic shear and cloud interactions are naturally included. These are some of the driving forces for the cloud internal turbulence \citep{DobbsEtAl2011} with many other driving mechanisms contributing \citep{FederrathEtAl2016,FederrathEtAl2017iaus}.

We note that magnetic fields were not included in the simulations. Magnetic fields are important for the structure and dynamics of the ISM \citep{PadoanNordlund2011,FederrathEtAl2011PRL,FederrathKlessen2012,KoertgenBanerjee2015,FogertyEtAl2016}, but instead of including all of the relevant detailed physics, we here focus on improving the aspect of initial and boundary conditions for cloud formation and dynamics. Follow-up simulations are need that include magnetic fields, detailed molecular chemistry and cooling, as well as star formation feedback, such as supernova explosions \citep{PadoanEtAl2016,PanEtAl2016,KoertgenBanerjee2016}. The strength of the present simulations is that they model a global galactic disc, thus providing better initial and boundary conditions for cloud formation than previous simulations.

\subsection{Cloud definition and tracking}
\label{sec:cloud_definition}

We identify clouds in our simulation using constant density contours {at $\rho_{\rm thresh} = 2.13\times 10^{22}\,\mathrm{g}\mathrm{cm}^{-3}$ ($n_\mathrm{H} \simeq 100\,\mathrm{cm}^{-3}$),} where the clouds are then the coherent structures within the contour line \citep[as in][]{FujimotoEtAl2014}. The threshold density for the cloud contour was selected based on the observed mean density of typical GMCs. Because the internal properties of the smaller clouds will be poorly resolved, we restrict our cloud analysis to clouds containing more than 1000~cells, giving an approximate 10~cells in each dimension, which is a the absolute minimum to capture the main fraction of rotational modes \citep{FederrathSurSchleicherBanerjeeKlessen2011,FederrathEtAl2014}. The exception to this is the resolution studies in Section~\ref{sec:resolution}, where differences in cell size result in too stringent a sub-selection of the cloud population, so we do not impose a minimum number of cells for the presentation of the resolution study below. At our two main analysis times of 60\,Myr and 240\,Myr, the total number of clouds in our highest-resolution simulation ($\Delta x=1\,\mathrm{pc}$) analysed is 1529 and 1313, respectively.

To follow the evolution of the clouds, we output the simulation data every $1\,\mathrm{Myr}$ and track the clouds between these data sets, assigning a unique tag number to the same structure at different times. If a cloud disappears during the evolution, we search for possible merger events and if unsuccessful, mark it as dispersed. A detailed description of the tracking algorithm can be found in \citet{TaskerTan2009}, although the cloud identification matches that in \citet{FujimotoEtAl2014}.

\subsection{Measuring the effective turbulence driving parameter}
\label{sec:b_parameter}

As described in the introduction, a key value for quantifying the cloud internal turbulence is the $b$ parameter, defined as the proportionality constant of the standard deviation--Mach number relation for non-magnetized and non-self-gravitating gas. For each cloud, 
\begin{eqnarray}
\sigma_{\rho/\rho_0} = b\mathcal{M},
\label{eq:b_parameter}
\end{eqnarray}
where $\sigma_{\rho/\rho_0}$ is the standard deviation of the gas density, $\rho$, within the cloud, normalised by the cloud mean density, $\rho_0$. The cloud's turbulent Mach number, $\mathcal{M} = \sigma_\mathrm{3D}/c_\mathrm{s}$, is given by the volume-weighted average of the local (cell-based) Mach number, where $\sigma_\mathrm{3D}$ and $c_\mathrm{s}$ are the three-dimensional velocity dispersion and sound speed, respectively. Finally, $b$ is the turbulence driving parameter \citep{FederrathKlessenSchmidt2008,FederrathDuvalKlessenSchmidtMacLow2010}, which we concentrate on measuring here.

This can be related to the gas density PDF if the latter is assumed to follow a log-normal distribution,
\begin{eqnarray}
{\rm PDF}(s) = \frac{1}{\sqrt{2 \pi \sigma_{s}^{2}}}\exp\left[-\frac{(s-\langle s \rangle)^2}{2\sigma_{s}^{2}}\right],
\label{eq:log_normal}
\end{eqnarray}
where $\sigma_s$ is the standard deviation of the logarithmic gas density with $s \equiv \ln(\rho/\rho_0)$. In which case, $b$ can be related to $\sigma_s$ with $\sigma_s^2 = \ln(1+b^2\mathcal{M}^2)$. However, note that this expression is dependent on the log-normal distribution of the gas, while Equation~\ref{eq:b_parameter} applies for any density PDF \citep{PadoanNordlundJones1997,NordlundPadoan1999,FederrathKlessenSchmidt2008,PriceFederrathBrunt2011}.

The turbulence driving parameter $b$ is determined by the mix of the solenoidal and compressive turbulent modes in the acceleration field that drives the turbulence. For fully solenoidal forcing (divergence-free), $b = 1/3$, while fully compressive forcing (curl-free) finds $b = 1$ in three-dimensional simulations \citep[]{FederrathKlessenSchmidt2008,FederrathDuvalKlessenSchmidtMacLow2010}. 

To estimate $b$ from the turbulent motions inside a cloud, we must first subtract the cloud's global motions \citep{FederrathSurSchleicherBanerjeeKlessen2011,PanEtAl2016}. The velocity dispersion is therefore calculated after the subtraction of the linear and angular momentum from the cloud motion. This is done by first subtracting the bulk velocity from the cloud's centre-of-mass. The cloud can then be regarded as a rigid body whose circular velocity about the core can be subtracted: $\mathbf{v}_\mathrm{circ} = \boldsymbol\omega \times \mathbf{r} = \mathbf{L} \times \mathbf{r} / (m r^2)$, where $\mathbf{r}$ is the position vector relative to the centre-of-mass, $\boldsymbol\omega$ is the angular velocity and $\mathbf{L}$ is the total angular momentum. The three-dimensional velocity dispersion about the cloud's centre-of-mass then becomes,
\begin{eqnarray}
\sigma_\mathrm{3D} &=& \sqrt{v_x^2 + v_y^2 + v_z^2},\quad\mathrm{with} \\
\label{eq:veldisp_3D}
v_x &=& v'_{x} - v_{cx} - v_{\mathrm{circ},x} \nonumber \\
v_y &=& v'_{y} - v_{cy} - v_{\mathrm{circ},y} \nonumber \\
v_z &=& v'_{z} - v_{cz} - v_{\mathrm{circ},z}, \nonumber
\end{eqnarray}
where $(v'_x, v'_y, v'_z)$ is the velocity of the gas, $(v_{cx},v_{cy},v_{cz})$ is the cloud's centre-of-mass velocity and $(v_{\mathrm{circ},x}, v_{\mathrm{circ},y}, v_{\mathrm{circ},z})$ is the circular velocity from the angular momentum about the cloud's centre. 

\section{Results}
\label{sec:results}

\subsection{Global and local gas structure}
\label{sec:gas_structure}

\begin{figure}
\includegraphics[width=1.0\linewidth]{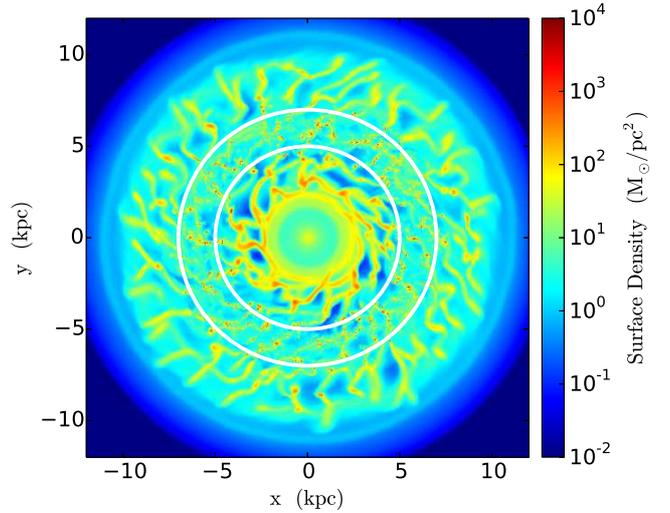}
\caption{Surface density of the whole galactic disc at $240\,\mathrm{Myr}$. 
The co-rotating, high-resolution region where we define the clouds is within the two white circles.}
\label{fig:figure1}
\end{figure}

The face-on view of the surface density in the whole galactic disc at $240\,\mathrm{Myr}$ (one orbital period at the outer edge of our analysis region at $r = 6.5$\,kpc.) is shown in Figure~\ref{fig:figure1}. The two white circles mark the inner and outer boundary of the high-resolution co-rotating region where we analyse the clouds. Using this technique we have a region with high resolution ($\Delta x_{\rm min} = 1\,\mathrm{pc}$) and at the same time include the global disc potential, with the appropriate boundary conditions for the high-resolution region. Since we do not drive a global spiral potential, the disc gas has a predominately flocculent structure, with the dense knots of cloud gas being connected by a filamentary warm ISM. The environment is clearly not quiescent, but a continuously interacting medley of collisions and tidal forces that shape the clouds.

\begin{figure}
\includegraphics[width=1.0\linewidth]{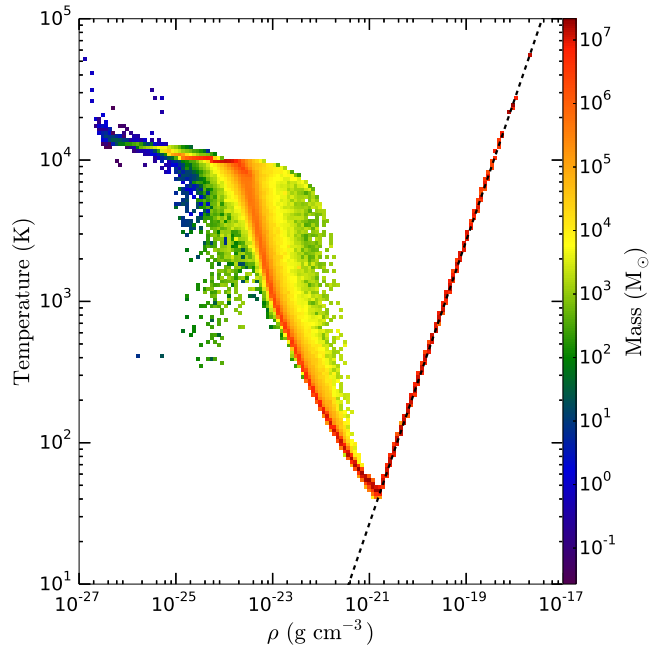}
\caption{Temperature vs.~density phase diagram showing the distribution of gas mass. The black dashed line corresponds to the pressure floor where gas follows a polytrope to suppress unresolved collapse.}
\label{fig:figure2}
\end{figure}

Figure~\ref{fig:figure2} describes the thermodynamical condition in the whole disc ISM via a two-dimensional density-temperature diagram of the distribution in the gas mass. Notably, there is no clear division in the gas between the hot, warm and cold ISM phases, but a continuous distribution connecting the cold clouds with the surrounding material. 

At densities $\rho \gtrsim 10^{21}\,\mathrm{g}\mathrm{cm}^{-3}$, the distribution becomes a steeply rising linear relation corresponding to the artificial pressure floor defined in Section~\ref{sec:code}. In this region, gas follows a polytrope with an adiabatic index of $\gamma = 2$, which is shown by the dashed black line. 
The pressure floor halts the collapse and prevents the formation of isolated cells with very high densities in which the Jeans length would otherwise not be resolved. While this regime was necessary to minimise numerical artifacts, it is clear that even on parsec scales, it is hard to resolve the molecular gas component of the clouds. Exactly where the simulations start to become resolution limited will be studied further in Section~\ref{sec:resolution}.

\subsection{Density PDF}
\label{sec:pdf}

\begin{figure}
\includegraphics[width=1.0\linewidth]{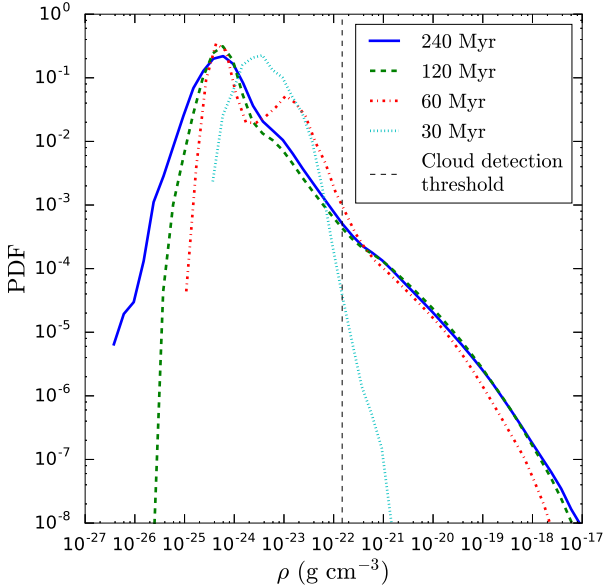}
\caption{Time evolution of the volume-weighted probability density function (PDF) of the logarithmic density ($\log_{10}\rho$) for the high-resolution region of the disc (between the white lines in Figure~\ref{fig:figure1}) with a height of ${\rm -100\,pc < z < 100\,pc}$.}
\label{fig:figure3}
\end{figure}

The evolution of the structure of the ISM can be explored using the density. Figure~\ref{fig:figure3} shows the volume-weighted density PDF for the high-resolution region of the galaxy marked between the two white circles in Figure~\ref{fig:figure1}. Four different simulation times are shown at 30\,Myr, 60\,Myr, 120\,Myr and 240\,Myr, with the additional vertical black dotted line showing the location of the cloud identification threshold, which sits at $\rho_{\rm thresh} = 2.13\times 10^{22}\,\mathrm{g}\mathrm{cm}^{-3}$.

At the earliest time of 30\,Myr, the galaxy disc is only partially fragmented. The PDF has a narrow profile that broadens as clouds form and begin to interact in the disc. From 60\,Myr to 240\,Myr, the profile continues to broaden but at a slower rate. Equation~\ref{eq:b_parameter} implies that the broadening of the PDF profile is due to an increase in the standard deviation of the logarithmic density. Previous research has suggested that this can occur due to a shift from solenoidal turbulence driving to compressive driving for the same Mach number \citep{FederrathKlessenSchmidt2008,PriceFederrathBrunt2011,PadoanNordlund2011,MolinaEtAl2012,KonstandinEtAl2012ApJ,NolanFederrathSutherland2015,FederrathBanerjee2015}. The evolution from a narrow to wide density PDF may therefore indicate a change from a largely solenoidal turbulence-dominated system to a more compressive regime as cloud and gas structures interact and collide. 

The cloud gas (right of the vertical dashed line) extends smoothly to higher densities, but with a profile that is not completely log-normal. While a log-normal distribution is expected from isothermal studies, deviations are frequently seen in other models due to shocks and other intermittent events \citep{FederrathDuvalKlessenSchmidtMacLow2010,GazolKim2013,NolanFederrathSutherland2015,FederrathBanerjee2015}. Moreover, our gas is not isothermal and the highest density end may be gravitationally collapsing, with support coming from the pressure floor ($\rho\gtrsim10^{-21}\,\mathrm{g}\,\mathrm{cm}^{-3}$) rather than internal turbulence.

\subsection{Cloud properties}
\label{sec:analysis}

\begin{figure*}
\centering
\includegraphics[width=0.8\linewidth]{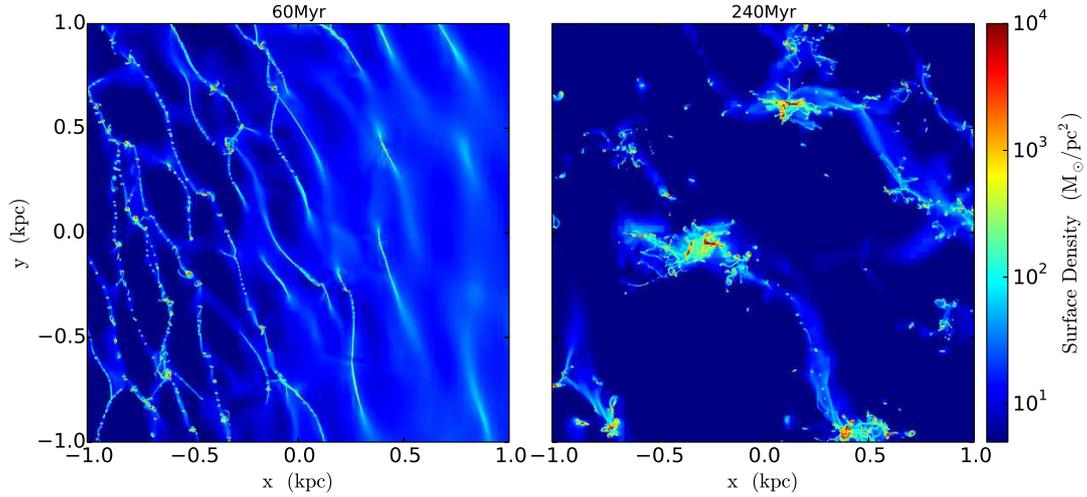}
\caption{Surface density images at $60 \,\mathrm{Myr}$ (left) and $240 \,\mathrm{Myr}$ (right). Images are $2\,\mathrm{kpc}$ across for the co-rotating region shown in Figure~\ref{fig:figure1} from $r = 5$ to $7\,\mathrm{kpc}$, where we define the clouds.}
\label{fig:figure4}
\end{figure*}

Clouds are formed in the galaxy disc through hydrodynamical and gravitational fragmentation. Figure~\ref{fig:figure4} shows a close-up of this process, displaying the surface density of the disc in a 2\,kpc region at early (60\,Myr) and late (240\,Myr) times. Both images are taken within the analysis region shown in Figure~\ref{fig:figure1}. The cloud environment changes substantially between these stages. At early times, the disc is in the process of fragmenting. Its rotational motion supports the gas against radial collapse, causing the initial structures to be tangential filaments. As the disc becomes more unstable, the rotation can no longer support the gas which collapses to create GMCs. As the dynamical time is shortest in the inner disc, the left-hand image in Figure~\ref{fig:figure3} shows this process occurring from right to left, with the outer gas on the right-hand side in an earlier state of collapse compared to the inner left-hand edge. By $240\,\mathrm{Myr}$, the disc has completely fragmented and interactions between the clouds now dominate the local dynamics.

\begin{figure*}
\centering
\includegraphics[width=0.8\linewidth]{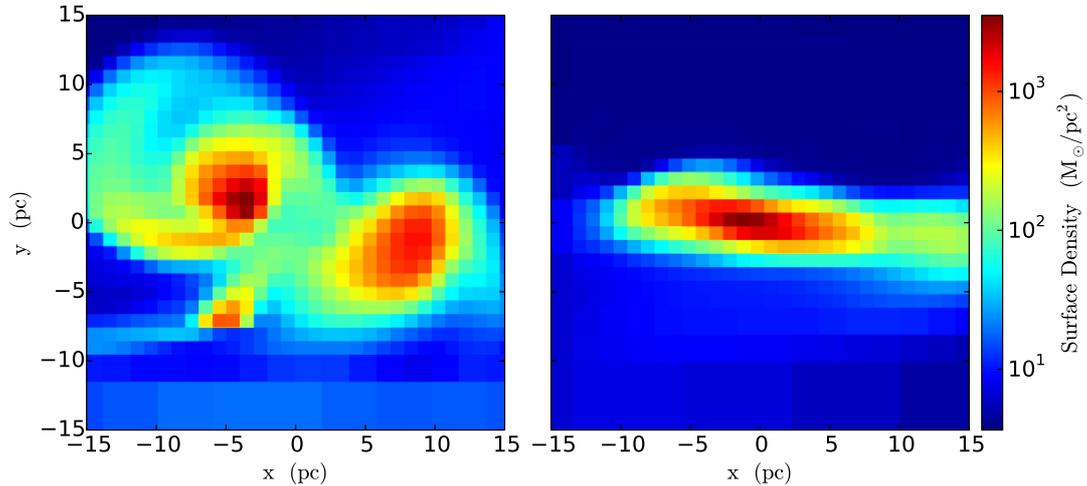}
\caption{Surface density of two typical clouds in the simulation at 240\,Myr. The left-hand cloud has a mass of $10^5$\,M$_\odot$, average radius of 9\,pc, velocity dispersion of 3\,km\,s$^{-1}$ and a virial parameter of 0.9. The right-hand cloud has a mass of $1.2\times 10^5$\,M$_\odot$, average radius 10\,kpc and also a velocity dispersion of 3\,km\,s$^{-1}$ and a virial parameter of 0.9.}
\label{fig:figure5}
\end{figure*}

\begin{figure*}
\centering
\includegraphics[width=0.8\linewidth]{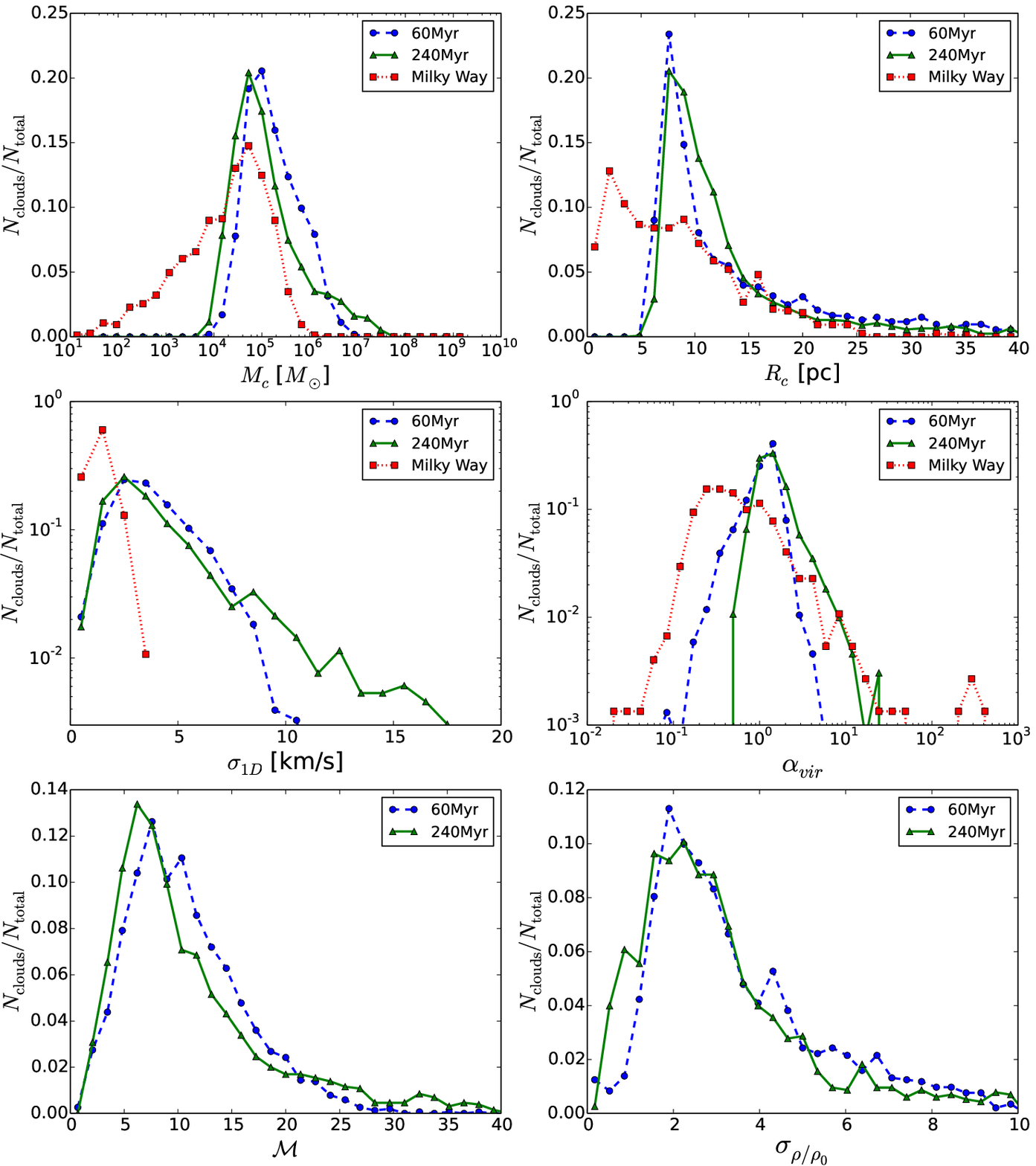}
\caption{Distributions of global cloud properties at $60$ and $240\,\mathrm{Myr}$ for clouds containing more than 1000~cells. Top left shows cloud mass, $M_\mathrm{c}$, top right shows cloud radius, $R_\mathrm{c}$, middle left shows 1D volume weighted velocity dispersion of clouds, $\sigma_\mathrm{1D}$, middle right shows virial parameter, $\alpha_\mathrm{vir}$, bottom left shows 3D volume weighted Mach number, $\mathcal{M}$, and bottom right shows the standard deviation, $\sigma_{\rho/\rho_0}$ of the density fluctuations. The red dotted line shows the distribution of clouds in the GRS Milky Way survey \citep{RomanDuvalEtAl2010}.}
\label{fig:figure6}
\end{figure*}

Close-ups of two typical clouds in the simulation are shown in Figure~\ref{fig:figure5}. The properties for these clouds match the peak values in the distributions shown in Figure~\ref{fig:figure6}. The left-hand cloud shows an interacting structure fragmented into three parts, while the right-hand cloud is a more quiescent object with a single core. Despite these visual differences, both clouds have similar mass, radii and velocity dispersions.

Figure~\ref{fig:figure6} turns to the properties of the clouds themselves, showing the distribution in their values at the early (60\,Myr in blue dashes) and late (240\,Myr in green solid) evolution times. As mentioned in Section~\ref{sec:cloud_definition}, only clouds with more than 1000~cells are included in this analysis. Overlaid with the red dotted line are the observational results from the GRS Milky Way survey from \citet{RomanDuvalEtAl2010}. The average and the standard deviation of cloud properties are summarised in Table~\ref{tab:table1}. 

\begin{table}
\def\arraystretch{1.0}
\setlength{\tabcolsep}{5.0pt}
	\centering
	\caption{Average and standard deviation of the cloud properties for clouds with more than 1000~cells (mass $M_\mathrm{c}$, radius $R_\mathrm{c}$, 1D velocity dispersion $\sigma_\mathrm{1D}$, virial parameter $\alpha_\mathrm{vir}$, Mach number $\mathcal{M}$ and density standard deviation $\sigma_{\rho/\rho_0}$) at $60$ and $240\,\mathrm{Myr}$.}
	\label{tab:table1}
	\begin{tabular}{cccccc} 
		\hline
		& \multicolumn{2}{|c|}{Average} & \multicolumn{2}{c|}{Standard deviation}\\
		\hline
		& $60\,\mathrm{Myr}$ & $240\,\mathrm{Myr}$ & $60\,\mathrm{Myr}$ & $240\,\mathrm{Myr}$\\
		\hline
		$M_\mathrm{c}$ ($\mathrm{M}_{\odot}$) & $4.1\times10^5$ & $8.9\times10^5$ & $7.2\times10^5$ & $30.9\times10^5$\\
		$R_\mathrm{c}$ (pc) & 15.4 & 15.4 & 11.8 & 13.2\\
		$\sigma_\mathrm{1D}$ ($\mathrm{km}\,\mathrm{s}^{-1}$) & 3.8 & 4.3 & 1.8 & 3.2\\
		$\alpha_{\rm vir}$ & 1.2 & 1.8 & 0.5 & 2.4\\
		$\mathcal{M}$ & 10.7 & 11.4 & 5.3 & 8.1\\
		$\sigma_{\rho/\rho_0}$ & 4.2 & 6.5 & 3.8 & 13.8\\
		\hline
	\end{tabular}
\end{table}

The top-left plot shows the distribution of the cloud mass. At 60\,Myr, the peak value sits at $M_\mathrm{c} \sim 10^5\,\mathrm{M}_{\odot}$. This has shifted slightly lower by 240\,Myr. The change is due to the clouds fragmenting into smaller structures as the disc finishes its period of instability. Both plots overlay the observed Milky Way profile, with approximately equivalent peak values. The simulation results have a high-mass tail due to the lack of star formation to remove the densest gas which increases over time. This creates a population of massive clouds whose bulk has built up via successive mergers and accretion. Its existence underscores the importance of interactions in the evolution of the clouds. The high-mass tail causes the standard deviation in cloud mass at $240\,\mathrm{Myr}$ ($30.9 \times 10^5\,\mathrm{M}_{\odot}$) to be significantly larger than that at $60\,\mathrm{Myr}$ ($7.2 \times 10^5\,\mathrm{M}_{\odot}$), as shown in Table~\ref{tab:table1}. Note that the mass here is plotted on a logarithmic scale (non-Gaussian distribution), indicating that the clouds in this later stage have a wide range of masses. The smaller clouds in the Milky Way data are not seen in the simulation plot, as we focus on our best-resolved population containing more than 1000~cells per cloud. 
 
The distribution in cloud radius is shown in the top right plot of Figure~\ref{fig:figure6}. We define the average radius of the cloud as
\begin{eqnarray}
\centering
R_\mathrm{c} = \sqrt{\frac{(A_{xy} + A_{yz} + A_{zx})}{3\pi}},
\label{eq:radius}
\end{eqnarray}
where $A_{xy}$ is the projected area of the cloud in the x-y plane, $A_{yz}$ in y-z plane and $A_{zx}$ in z-x plane.

The peak cloud radius remains the same between 60\,Myr and 240\,Myr, but the profile becomes broader at later times. The average radius remains the same, with $\langle R_\mathrm{c} \rangle = 15.4\,\mathrm{pc}$. However, Figure~\ref{fig:figure4} suggests that the cloud shape will be changing more considerably as it goes from being dominated by a fragmenting filament to one controlled both by the warm ISM filaments and cloud interactions. The range in values at both times are consistent with those measured in the Milky Way, but once again we omit the smaller clouds from our analysis. 

The middle-left panel shows the distribution of the one-dimensional volume-weighted velocity dispersion after the subtraction of the rotational velocity. The evolution shows a broadening in the range of values, with a late population of high velocity dispersion clouds. These likely correspond to the high-mass tail seen in the mass distribution, whose increased gravitational potential from successive mergers boosts the velocity dispersion. While this larger population is not seen in the Milky Way the peak value at both 60\,Myr and 240\,Myr is close to that seen observationally at about $2$--$3\,\mathrm{km}\,\mathrm{s}^{-1}$.

The virial parameter, $\alpha_\mathrm{vir}$, describes the ratio of kinetic to gravitational energy using observational parameters. The value is defined as
\begin{eqnarray} 
\alpha_\mathrm{vir} = \frac{5(c_\mathrm{s}^2 + \sigma_\mathrm{1D}^2)R_\mathrm{c}}{GM_\mathrm{c}},
\label{eq:virial}
\end{eqnarray}
where $c_\mathrm{s}^2=\gamma k_\mathrm{B} T / (\mu m_H)$, $T$ is the volume-weighted average cloud temperature, $\gamma = 1.67$ and $\mu = 2.3$, the molecular weight in a GMC. The 1D velocity dispersion is $\sigma_\mathrm{1D} = \frac{1}{\sqrt{3}} \sqrt{v_x^2 + v_y^2 + v_z^2}$, where the velocity components $v_x$, $v_y$, and $v_z$ were corrected for linear and angular momentum contributions according Equation~\ref{eq:veldisp_3D}, $R_\mathrm{c}$ is the cloud radius and $M_\mathrm{c}$ is the cloud mass. A value of $\alpha_\mathrm{vir} > 1$ implies that kinetic energy is dominant, while $\alpha_\mathrm{vir} < 1$ indicates that gravity is dominant and therefore, the cloud is gravitationally bound \citep[]{BertoldiMcKee1992}.

The virial parameter distribution is shown in the middle right panel of Figure~\ref{fig:figure6}. The averages given in Table~\ref{tab:table1} show $\langle \alpha_\mathrm{vir} \rangle = 1.2$ at $60\,\mathrm{Myr}$ and $\langle \alpha_\mathrm{vir} \rangle = 1.8$ at $240\,\mathrm{Myr}$. At later times, the virial parameter average shifts to a slightly higher value. However, the distribution shows that the peak value remains unchanged at just greater than unity, suggesting most clouds are borderline bound throughout their lives. At later times, a population of less-bound clouds appears, suggesting that the products of successive mergers and interactions tend to not be gravitationally bound. The Milky Way data likewise finds that clouds sit on the brink of being bound, with the data showing a broad range of values. There are clearly differences between the simulated and observed distributions, but given the uncertainties in both the simulations and the observations, we do not expect to find significantly better agreement.

Left on the bottom row shows the distribution in the cloud Mach number, $\mathcal{M} = \sigma_\mathrm{3D}/c_\mathrm{s}$, while the bottom right is the density standard deviation. These are the two properties that control the value of the turbulence parameter, $b$, which we will focus on in the next subsection. Both distributions have a minor dependence on time. The peak value of Mach number at $240\,\mathrm{Myr}$ is $\mathcal{M} \sim 6.0$, which is slightly lower than $\mathcal{M} \sim 8.0$ at $60\,\mathrm{Myr}$. This comes from the broader range in velocity dispersion at later times that we saw in the plot above; the broadening at 240\,Myr causes the velocity dispersion to peak at a lower value. A similar but weaker trend can be seen in the distribution of the standard deviation, $\sigma_{\rho/\rho_0}$; namely clouds in the later stage generate lower $\sigma_{\rho/\rho_0}$ than early on in the simulations. The distribution is also slightly broader, with clouds at 240\,Myr showing a more varied internal density range than at 60\,Myr. This likely stems directly from each cloud's history, determined by interactions, birth location and the local environment. The average values for the standard deviation in Table~\ref{tab:table1} are high due to being dominated by 1\% of outlying clouds that are undergoing gravitational collapse. 

\subsection{Turbulence driving mode}
\label{sec:drivingmode}

\begin{figure}
\centering
\includegraphics[width=1.0\linewidth]{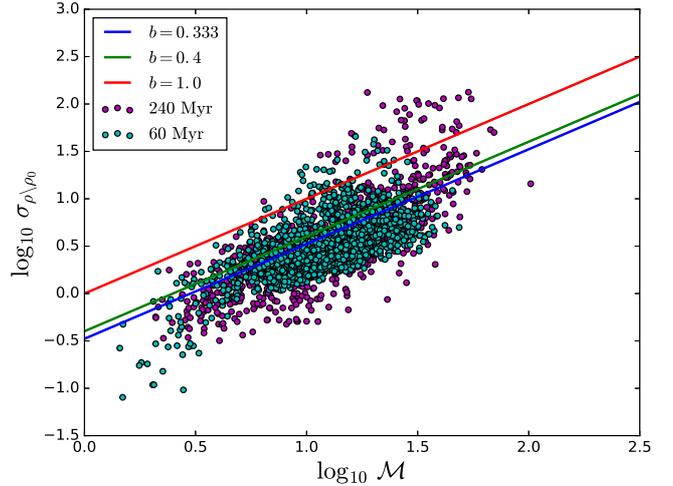}
\caption{The relation between cloud Mach number $\mathcal{M}$ and the standard deviation of density fluctuations $\sigma_{\rho/\rho_0}$. Teal points show individual clouds at $60\,\mathrm{Myr}$ and purple points are clouds at $240\,\mathrm{Myr}$. The lines show the gradient corresponding to different values of $b$, based on Equation~\ref{eq:b_parameter}.}
\label{fig:figure7}
\end{figure}

One of the most important results in our study is shown in Figure~\ref{fig:figure7}. It compares the standard deviation $\sigma_{\rho/\rho_0}$--Mach number $\mathcal{M}$ relation for the clouds at the early $60$\,Myr simulation time and those at $240\,\mathrm{Myr}$ to investigate the change in turbulence modes over the course of the simulation. As with Figure~\ref{fig:figure6}, we only plot clouds with more than 1000~cells. The purple points correspond to clouds at $240\,\mathrm{Myr}$ and the teal points for $60\,\mathrm{Myr}$, with the red line showing the gradient for $b = 1.0$ (pure compressive driving) in Equation~\ref{eq:b_parameter} and the green and blue lines marking $b = 0.4$ and $b = 0.3$ (pure solenoidal driving), respectively. 

Surprisingly, despite the significant difference in disc structure at the 60\,Myr and 240\,Myr times shown in Figure~\ref{fig:figure4}, the best fit for the cloud turbulence modes is $b\sim0.3$ for both times. The reason for this agreement can be seen from the last panels in Figure~\ref{fig:figure6}, where the peaks in both the Mach number and standard deviation $\sigma_{\rho/\rho_0}$ shift slightly towards lower values with time, compensating for one another to give the same average $b$. Physically speaking, this supports the picture that the clouds break into smaller clouds during the disc gravitational fragmentation and subsequent cloud interactions. While some clouds merge to form the high-end tail of the mass distribution, a substantial fraction of clouds have a lower Mach number and density contrast post-fragmentation, which balance to give a similar mix of turbulence modes. This average value of $b$ indicates that the clouds in the galaxy simulation are typically dominated by solenoidal turbulence driving. However, in the next section we will note this is a strictly lower limit, because of the limited numerical resolution currently achievable in these simulations of disc galaxies.

\begin{figure}
\centering
\includegraphics[width=1.0\linewidth]{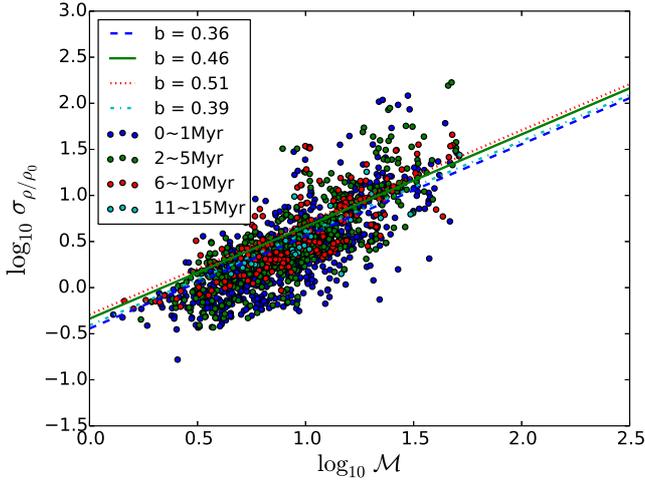}
\caption{Distribution of $b$ for cloud populations at different ages. The clouds are tracked every $1\,\mathrm{Myr}$ from $220$ to $240\,\mathrm{Myr}$ and then grouped into populations based on the time since the cloud formed. The age of clouds with blue data points is $0$--$1\,\mathrm{Myr}$, green is $2$--$5\,\mathrm{Myr}$, red is $6$--$10\,\mathrm{Myr}$ and cyan cloud points are for clouds with age $11$--$15\,\mathrm{Myr}$. The lines show different values of $b$, with the blue-dashed, green-solid, red-dotted and cyan-dot-dash showing the best fit for $b$ in each of the cloud age data sets with the same colour.}
\label{fig:figure8}
\end{figure}

More important than the average value of $b$ (which is not converged with resolution), we find a very wide distribution of $b$, with values covering the full range from purely solenoidal ($b\sim1/3$) to purely compressive driving ($b\sim1$) and mixed values in between. Thus, the turbulence driving parameter for individual clouds is far from constant. This shows that $b$ should be considered a varying parameter that changes between individual clouds. The range of $b$ values in the cloud population increases with simulation time, as the distribution at 240\,Myr in Figure~\ref{fig:figure7} is somewhat broader than at 60\,Myr. This suggests that a cloud's history or its environment might be controlling the evolution of $b$. To test the idea that clouds are born with a constant $b$ and interactions drive more compressive or solenoidal driving, we plot the same relation as in Figure~\ref{fig:figure7} for clouds of different ages. This is shown in Figure~\ref{fig:figure8}.

As described in Section~\ref{sec:cloud_definition}, we followed the evolution of $b$ for individual clouds by tracking them between simulation outputs. We did this for clouds between 220--240\,Myrs, during the fully fragmented phase of the disc where cloud interactions dominated their evolution. In Figure~\ref{fig:figure8}, clouds that are between $0$--$1$\,Myr old are marked in dark blue, clouds between $2$--$5$\,Myr old are in green, $6$--$10$\,Myr old clouds are in red and $11$--$15$\,Myr old clouds are cyan points. Lines of the same colour show the best-fit $b$ parameter for each population. There is initially a small amount of evolution in $b$ as the cloud ages, moving from an average $b = 0.36$ for the youngest clouds towards an average $b = 0.51$ for clouds between $6$--$10$\,Myr. However, this number decreases again slightly for the oldest clouds. While not major changes, these are likely to be due to the changes in compressive modes as the cloud undergoes interactions during its life. However, this effect is clearly weak, suggesting no strong relation between the turbulence modes and cloud mergers or lifetime. We also attempted to find a stronger correlation by examining the values of $b$ before and after individual recorded mergers. No emerging trend was seen, with $b$ staying approximately constant during the interaction. It appears instead that clouds overall retain the $b$ they are formed with, although this value can take on a wide range of numbers.  

\subsection{The compressive ratio}

\begin{figure}
\centering
\includegraphics[width=1.0\linewidth]{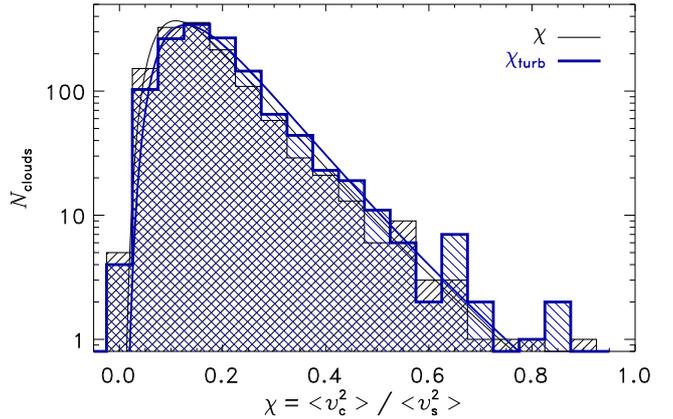}
\caption{Histograms of the total compressive ratio, $\chi$, (thin line histogram) and its turbulent component, $\chi_\mathrm{turb}$ (thick line histogram). The mean and standard deviation of these distributions are $\chi=0.20\pm0.13$ and $\chi_\mathrm{turb}=0.22\pm0.14$. The thin and thick curves are log-normal fits to the distributions as in \citet{PanEtAl2016}.}
\label{fig:figure9}
\end{figure}

Our main goal here is to quantify the relative strength of solenoidal to compressive modes in the turbulence driving of molecular clouds formed by galactic dynamics and cloud-cloud collisions. An important diagnostic of this is the \emph{compressive ratio} \citep{KritsukEtAl2007,FederrathDuvalKlessenSchmidtMacLow2010,FederrathEtAl2011PRL,PadoanEtAl2016,PanEtAl2016}, which quantifies the fraction of compressive modes in the turbulent velocity field. Here we follow the recent definition of the compressive ratio proposed in \citet{PanEtAl2016},
\begin{equation} \label{eq:compratio}
\chi = \langle v_\mathrm{c}^2\rangle / \langle v_\mathrm{s}^2\rangle,
\end{equation}
where $v_\mathrm{c}$ and $v_\mathrm{s}$ are the compressive and solenoidal components of the velocity field, respectively. The two velocity components are derived with the Helmholtz decomposition of the velocity field in Fourier space \citep[e.g.,][]{FederrathKlessenSchmidt2009,FederrathDuvalKlessenSchmidtMacLow2010,FederrathKlessen2013}. In addition to the total compressive ratio $\chi$ given by Equation~(\ref{eq:compratio}), we consider a purely turbulent version, $\chi_\mathrm{turb}$, where we subtracted the centre-of-mass velocity as well as the mean velocities $v_r$, $v_\theta$ and $v_\phi$ in spherical shells centred on the centre of mass of each cloud in spherical coordinates ($r$,$\theta$,$\phi$) \citep{FederrathSurSchleicherBanerjeeKlessen2011,SurEtAl2012}. This removes systematic motions from the velocity field such that only turbulent fluctuations remain. This method is similar to the method used in \citet{PanEtAl2016} to remove systematic motions and to isolate the turbulence in the clouds.

Figure~\ref{fig:figure9} shows the frequency distribution of the total compressive ratio $\chi$ and the turbulent compressive ratio $\chi_\mathrm{turb}$ of all the clouds with at least 1000 cells in our sample. Similar to the distributions of the turbulent driving parameter $b$, we find broad distributions of the compressive ratio, with values of $\chi$ covering a wide range from purely solenoidal ($\chi=0$) to strongly compressive ($\chi\sim1$). The mean and standard deviation of these distributions are $\chi=0.20\pm0.13$ and $\chi_\mathrm{turb}=0.22\pm0.14$, suggesting that most clouds are dominated by solenoidal modes. This is similar to the conclusion drawn from the sample of clouds in \citet{PanEtAl2016}, who find broad distributions and typical values of $\chi\sim0.3$ from simulations of supernova driving. Our values of $\chi$ are slightly lower, which is expected for turbulence purely driven by galactic dynamics (which drives solenoidal motions due to galactic shear) compared to turbulence driven by supernova explosions (which drive more compressive modes compared to galactic shear) \citep{FederrathEtAl2016,FederrathEtAl2017iaus}.

In agreement with \citet{PanEtAl2016} we do not find a significant difference between the total and turbulent compressive ratio. The most likely explanation for the similarity of $\chi$ and $\chi_\mathrm{turb}$ is that for $\chi_\mathrm{turb}$ one removes systematic compressive motions (infall) from the numerator $\langle v_\mathrm{c}\rangle$ and at the same time removes systematic solenoidal motions (large-scale rotation) from the denominator $\langle v_\mathrm{s}\rangle$, such that the ratio of the two nearly stays the same.

In summary, we find broad distributions of the compressive ratio of our cloud sample formed by galactic dynamics and cloud-cloud collisions, with typical values of $\chi\sim0.2$, i.e., dominated by solenoidal modes. Following the suggestions by \citet{PanEtAl2016}, this value of $\chi$ may be related to the turbulence driving parameter $b\sim[\chi/(1+\chi)]^{1/2}\sim0.4$, which is broadly consistent with the average value of $b$ found directly in Section~\ref{sec:drivingmode}.

\subsection{Resolution study}
\label{sec:resolution}

\begin{figure}
\centering
\includegraphics[width=1.0\linewidth]{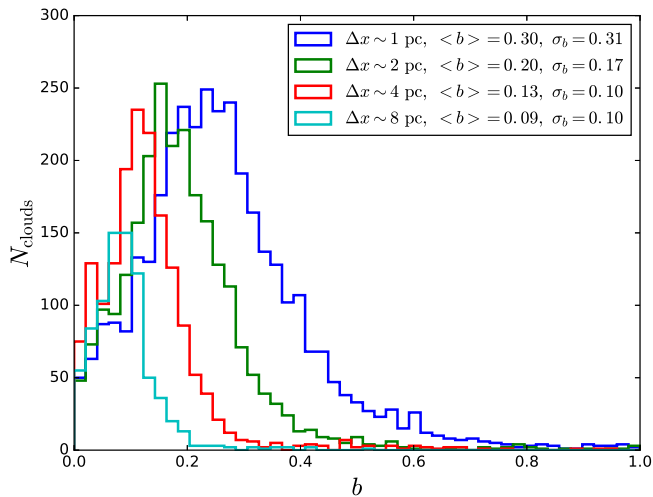}
\caption{The variation of the turbulence driving parameter, $b$, with resolution. We find that the average $b$ value increases systematically with increasing resolution, which means that our measurements represent lower limits. However, every resolution shows that $b$ varies by factors of a few within the cloud population.}
\label{fig:figure10}
\end{figure}

\begin{figure*}
\includegraphics[width=1.0\linewidth]{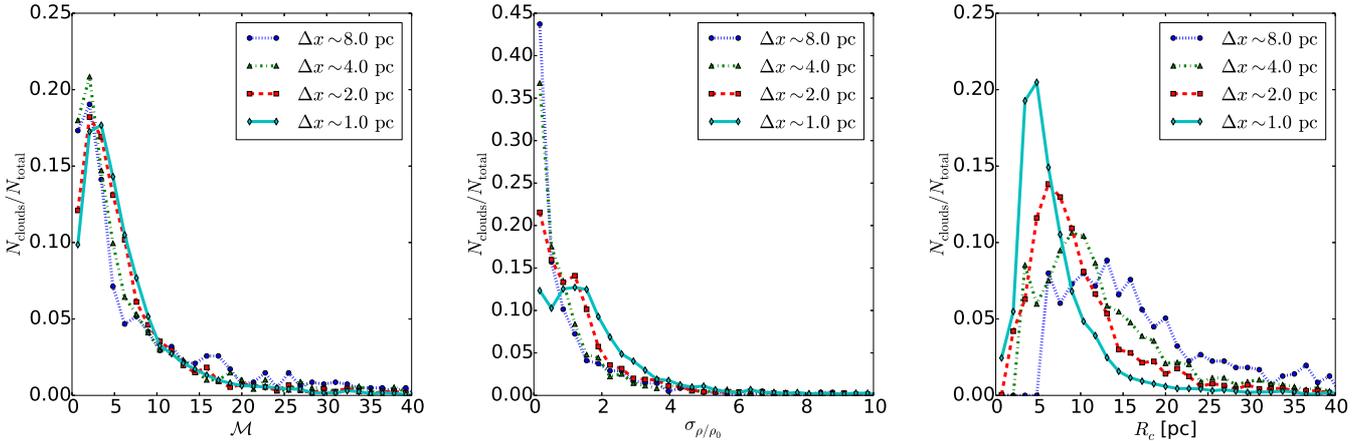}
\caption{Distributions for the 3D volume-weighted Mach number, $\mathcal{M}$ (left), the density standard deviation, $\sigma_{\rho/\rho_0}$ (middle), and the cloud radius, $R_\mathrm{c}$ (right) for increasing maximum numerical resolution between $8 > \Delta x > 1$\,pc, where $\Delta x$ is the smallest cell size. The blue circles-dots line shows the distribution for $\Delta x = 8\,\mathrm{pc}$, the green triangle-dashed line for $\Delta x = 4\,\mathrm{pc}$, red square-long dashed line for $\Delta x = 2\,\mathrm{pc}$ and cyan diamond-solid line for $\Delta x = 1\,\mathrm{pc}$. The resolution dependence of our $b$ value is primarily caused by the resolution dependence in $\sigma_{\rho/\rho_0}$, while the Mach number is converged with numerical resolution of $\Delta x \lesssim 4\,\mathrm{pc}$.}
\label{fig:figure11}
\end{figure*}

One important consideration in our quantification of the turbulence driving parameter $b$ is the effect of numerical resolution of the simulations. While the simulation refines down to cells with sizes $\Delta x = 1\,\mathrm{pc}$, the internal cloud turbulence is a challenging property to resolve and converge on. To test the robustness of our quantitative values, we compared simulations at four different refinement levels, corresponding to limiting resolutions (smallest cell size) of $\Delta x = 8\,\mathrm{pc}$, $\Delta x = 4\,\mathrm{pc}$, $\Delta x = 2\,\mathrm{pc}$ and our main simulation resolution at $\Delta x = 1\,\mathrm{pc}$. The resulting values for the cloud parameter $b$ at each resolution are shown in Figure~\ref{fig:figure10}. The dark blue line shows the distribution for our main simulation run, while green, red and cyan mark the distributions for simulations of decreasing resolution. The legend shows the average $b$ value with its standard deviation. Due to the large size of the cells in the lower-resolution run, Figures~\ref{fig:figure10} and accompanying Figure~\ref{fig:figure11} show all clouds in the simulation, not only those with cell number above~1000.

Our resolution test shows a systematic dependence of $b$ on $\Delta x$, with the turbulence value increasing at higher resolutions (decreasing $\Delta x$). In our lowest resolution simulation with $\Delta x = 8\,\mathrm{pc}$ (cyan line), the average $b$ for all clouds is 0.09. This increases to $0.3$ for a limiting resolution of $\Delta x = 1\,\mathrm{pc}$, showing an increase in the resolved compressive modes. Such a trend means that we cannot determine an absolute value for $b$ in our global models and the average value of $b\sim0.3$ should be considered a lower limit even in the highest-resolution simulations. The standard deviation in the values for $b$ also increases with resolution, although this is likely to be due to the wider range in cloud sizes that the smaller cells allow, as we will investigate next.

The origin of the resolution dependence of $b$ can be seen in Figure~\ref{fig:figure11}. The three panels show how the profiles for the 3D volume-weighted Mach number, standard deviation $\sigma_{\rho/\rho_0}$ and the cloud radius vary with resolution. The Mach number profile shows a nearly constant distribution shape past our lowest resolution level, suggesting that this property has converged in the simulations. However, both the standard deviation of density fluctuations and the radius of the clouds shift to systematically higher and lower values, respectively, with increasing resolution. Full convergence is likely to be achieved when the simulation resolves the sonic scale ($\sim 0.1$\,pc), where the turbulence transitions from supersonic to subsonic \citep[]{VazquezBallesterosKlessen2003,FederrathDuvalKlessenSchmidtMacLow2010,Hopkins2013PDF}. This is because below the sonic scale, the turbulent density fluctuations become small compared to the large-scale supersonic fluctuations. The sonic scale is also the scale on which filaments and dense cores are expected to form \citep[]{GoodmanEtAl1998,ArzoumanianEtAl2011,AndreEtAl2014,Federrath2016}. Unfortunately, this is still out of reach for our global galaxy simulations. Future galaxy simulations, however, will be able to reach down to sub-pc scales.

For the main result of this work, we emphasise that despite the resolution dependence of the exact value of $b$, all resolutions show that the cloud population has a wide distribution of $b$ values, indicating that galaxy dynamics and cloud-cloud collisions alone (without feedback from star formation) can produce the full range of cloud driving parameters, from purely solenoidal driving ($b\sim1/3$) to purely compressive driving ($b\sim1$). We thus expect that these clouds can have a wide range of star formation activity, as star formation depends critically on the driving and statistics of the turbulence \citep{FederrathKlessen2012,PadoanEtAl2014,FederrathEtAl2016}.

\section{Conclusions}
\label{sec:conclusions}

We performed a three-dimensional hydrodynamical simulation of a global isolated disc galaxy and explored the turbulence properties in the giant molecular cloud population formed through self-gravity and cloud interactions. The turbulent modes were quantitatively explored by using the turbulence driving parameter $b$, which is around $0.3$ for solenoidal modes and $1$ for purely compressive driving of the turbulence. Our main results are as follows.

\begin{enumerate}

\item In the early and the late stages of galaxy disc evolution, the average $b$ for the cloud population is almost the same at $b \sim 0.3$. This is despite the morphology of the disc structure being quite different, due to only being partially fragmented at the earlier times. Our subsequent resolution study indicated that this average value of $b$ is a strictly lower limit on the true value and resolution down to $\sim0.1$\,pc is likely required to achieve convergence.
 
\item The distribution of $b$ exhibits wide scatter through the cloud population with $b$ varying by factors of a few. This means that each cloud can have a significantly different turbulence driving mode, ranging from purely solenoidal, divergence-free driving ($b\sim 0.3$) to purely compressive, curl-free driving ($b\sim 1$).

\item The main result of a wide distribution of turbulence properties of the clouds is supported by calculations of the compressive ratio; the ratio of compressive to solenoidal modes in the turbulent velocity field, $\chi$. Values of $\chi$ likewise took on a wide range, with the mean value of $\chi \sim 0.2$, suggesting primarily solenoidal modes on average, but with individual clouds having values reaching far into the highly compressible regime.

 \item We tracked the clouds through the simulation to compare changes during the cloud lifetime. Clouds live in a highly dynamic environment with multiple interactions occurring over the cloud's life. Despite this, we find that the $b$ distribution does not depend significantly on the cloud age.

\end{enumerate}

We conclude that the turbulent modes within GMCs do not significantly depend on the evolutionary stage of the galaxy disc, nor on the merger history of a cloud. However, there is always a wide distribution of modes found within the cloud population. This means that a constant $b$ cannot be selected, but must instead be treated as a varying parameter when looking at individual clouds.

\section*{Acknowledgements}
We thank the anonymous referee for their detailed comments, which helped to improve this work. Numerical computations were carried out on Cray XC30 at the Center for Computational Astrophysics (CfCA) of the National Astronomical Observatory of Japan. The authors would like to thank the YT development team for many helpful analysis support \citep{TurkEtAl2011}.
C.F.~acknowledges funding provided by the Australian Research Council's Discovery Projects (grants~DP150104329 and~DP17010603).
C.F.~further thanks the Leibniz Rechenzentrum and the Gauss Centre for Supercomputing (grants pr32lo, pr48pi and GCS Large-scale project 10391), the Partnership for Advanced Computing in Europe (PRACE grant pr89mu), and the Australian National Computational Infrastructure (grant ek9), as well as the Pawsey Supercomputing Centre with funding from the Australian Government and the Government of Western Australia. E.J.T.~was partially funded by the MEXT grant for the Tenure Track System. D.M.S.~gratefully acknowledges funding from the Australian Government's New Colombo Plan. This work is partly supported by JSPS Grant-in-Aid for Scientific Research  Number 15K0514.


\begin{thebibliography}{79}
\expandafter\ifx\csname natexlab\endcsname\relax\def\natexlab#1{#1}\fi

\bibitem[{{Abel} {et~al.}(1997){Abel}, {Anninos}, {Zhang}, \&
  {Norman}}]{Abel1997}
{Abel}, T., {Anninos}, P., {Zhang}, Y., \& {Norman}, M.~L. 1997, \na, 2, 181

\bibitem[{{Andr{\'e}} {et~al.}(2014){Andr{\'e}}, {Di Francesco},
  {Ward-Thompson}, {Inutsuka}, {Pudritz}, \& {Pineda}}]{AndreEtAl2014}
{Andr{\'e}}, P., {Di Francesco}, J., {Ward-Thompson}, D., {et~al.} 2014,
  Protostars and Planets VI, 27

\bibitem[{{Arzoumanian} {et~al.}(2011){Arzoumanian}, {Andr{\'e}}, {Didelon},
  {K{\"o}nyves}, {Schneider}, {Men'shchikov}, {Sousbie}, {Zavagno}, {Bontemps},
  {di Francesco}, {Griffin}, {Hennemann}, {Hill}, {Kirk}, {Martin}, {Minier},
  {Molinari}, {Motte}, {Peretto}, {Pezzuto}, {Spinoglio}, {Ward-Thompson},
  {White}, \& {Wilson}}]{ArzoumanianEtAl2011}
{Arzoumanian}, D., {Andr{\'e}}, P., {Didelon}, P., {et~al.} 2011, \aap, 529, L6

\bibitem[{{Benincasa} {et~al.}(2013){Benincasa}, {Tasker}, {Pudritz}, \&
  {Wadsley}}]{BenincasaEtAl2013}
{Benincasa}, S.~M., {Tasker}, E.~J., {Pudritz}, R.~E., \& {Wadsley}, J. 2013,
  \apj, 776, 23

\bibitem[{{Bertoldi} \& {McKee}(1992)}]{BertoldiMcKee1992}
{Bertoldi}, F., \& {McKee}, C.~F. 1992, \apj, 395, 140

\bibitem[{{Brunt}(2010)}]{Brunt2010}
{Brunt}, C.~M. 2010, \aap, 513, A67

\bibitem[{{Brunt} {et~al.}(2009){Brunt}, {Heyer}, \& {Mac
  Low}}]{BruntHeyerMacLow2009}
{Brunt}, C.~M., {Heyer}, M.~H., \& {Mac Low}, M. 2009, \aap, 504, 883

\bibitem[{{Bryan}(1999)}]{Bryan1999}
{Bryan}, G.~L. 1999, Comput.~Sci.~Eng., Vol.~1, No.~2, p.~46 - 53, 1, 46

\bibitem[{{Bryan} \& {Norman}(1997)}]{BryanNorman1997}
{Bryan}, G.~L., \& {Norman}, M.~L. 1997, in Astronomical Society of the Pacific
  Conference Series, Vol. 123, Computational Astrophysics; 12th Kingston
  Meeting on Theoretical Astrophysics, ed. D.~A. {Clarke} \& M.~J. {West}, 363

\bibitem[{{Bryan} {et~al.}(2014){Bryan}, {Norman}, {O'Shea}, {Abel}, {Wise},
  {Turk}, {Reynolds}, {Collins}, {Wang}, {Skillman}, {Smith}, {Harkness},
  {Bordner}, {Kim}, {Kuhlen}, {Xu}, {Goldbaum}, {Hummels}, {Kritsuk}, {Tasker},
  {Skory}, {Simpson}, {Hahn}, {Oishi}, {So}, {Zhao}, {Cen}, {Li}, \& {Enzo
  Collaboration}}]{BryanEtAl2014}
{Bryan}, G.~L., {Norman}, M.~L., {O'Shea}, B.~W., {et~al.} 2014, \apjs, 211, 19

\bibitem[{{Dobbs} {et~al.}(2011){Dobbs}, {Burkert}, \&
  {Pringle}}]{DobbsEtAl2011}
{Dobbs}, C.~L., {Burkert}, A., \& {Pringle}, J.~E. 2011, \mnras, 413, 2935

\bibitem[{{Elmegreen} \& {Scalo}(2004)}]{ElmegreenScalo2004}
{Elmegreen}, B.~G., \& {Scalo}, J. 2004, \araa, 42, 211

\bibitem[{{Federrath}(2013{\natexlab{a}})}]{Federrath2013}
{Federrath}, C. 2013{\natexlab{a}}, \mnras, 436, 1245

\bibitem[{{Federrath}(2013{\natexlab{b}})}]{Federrath2013sflaw}
---. 2013{\natexlab{b}}, \mnras, 436, 3167

\bibitem[{{Federrath}(2016)}]{Federrath2016}
---. 2016, \mnras, 457, 375

\bibitem[{{Federrath} \& {Banerjee}(2015)}]{FederrathBanerjee2015}
{Federrath}, C., \& {Banerjee}, S. 2015, \mnras, 448, 3297

\bibitem[{{Federrath} {et~al.}(2011{\natexlab{a}}){Federrath}, {Chabrier},
  {Schober}, {Banerjee}, {Klessen}, \& {Schleicher}}]{FederrathEtAl2011PRL}
{Federrath}, C., {Chabrier}, G., {Schober}, J., {et~al.} 2011{\natexlab{a}},
  PhRvL, 107, 114504

\bibitem[{{Federrath} \& {Klessen}(2012)}]{FederrathKlessen2012}
{Federrath}, C., \& {Klessen}, R.~S. 2012, \apj, 761, 156

\bibitem[{{Federrath} \& {Klessen}(2013)}]{FederrathKlessen2013}
---. 2013, \apj, 763, 51

\bibitem[{{Federrath} {et~al.}(2008){Federrath}, {Klessen}, \&
  {Schmidt}}]{FederrathKlessenSchmidt2008}
{Federrath}, C., {Klessen}, R.~S., \& {Schmidt}, W. 2008, \apjl, 688, L79

\bibitem[{{Federrath} {et~al.}(2009){Federrath}, {Klessen}, \&
  {Schmidt}}]{FederrathKlessenSchmidt2009}
---. 2009, \apj, 692, 364

\bibitem[{{Federrath} {et~al.}(2010){Federrath}, {Roman-Duval}, {Klessen},
  {Schmidt}, \& {Mac Low}}]{FederrathDuvalKlessenSchmidtMacLow2010}
{Federrath}, C., {Roman-Duval}, J., {Klessen}, R.~S., {Schmidt}, W., \& {Mac
  Low}, M. 2010, \aap, 512, A81

\bibitem[{{Federrath} {et~al.}(2014){Federrath}, {Schr{\"o}n}, {Banerjee}, \&
  {Klessen}}]{FederrathEtAl2014}
{Federrath}, C., {Schr{\"o}n}, M., {Banerjee}, R., \& {Klessen}, R.~S. 2014,
  \apj, 790, 128

\bibitem[{{Federrath} {et~al.}(2011{\natexlab{b}}){Federrath}, {Sur},
  {Schleicher}, {Banerjee}, \&
  {Klessen}}]{FederrathSurSchleicherBanerjeeKlessen2011}
{Federrath}, C., {Sur}, S., {Schleicher}, D.~R.~G., {Banerjee}, R., \&
  {Klessen}, R.~S. 2011{\natexlab{b}}, \apj, 731, 62

\bibitem[{{Federrath} {et~al.}(2016){Federrath}, {Rathborne}, {Longmore},
  {Kruijssen}, {Bally}, {Contreras}, {Crocker}, {Garay}, {Jackson}, {Testi}, \&
  {Walsh}}]{FederrathEtAl2016}
{Federrath}, C., {Rathborne}, J.~M., {Longmore}, S.~N., {et~al.} 2016, \apj,
  832, 143

\bibitem[{{Federrath} {et~al.}(2017){Federrath}, {Rathborne}, {Longmore},
  {Kruijssen}, {Bally}, {Contreras}, {Crocker}, {Garay}, {Jackson}, {Testi}, \&
  {Walsh}}]{FederrathEtAl2017iaus}
{Federrath}, C., {Rathborne}, J.~M., {Longmore}, S.~N., {et~al.} 2017, in IAU
  Symposium, Vol. 322, IAU Symposium, ed. R.~M. {Crocker}, S.~N. {Longmore}, \&
  G.~V. {Bicknell}, 123--128

\bibitem[{{Ferland} {et~al.}(1998){Ferland}, {Korista}, {Verner}, {Ferguson},
  {Kingdon}, \& {Verner}}]{FerlandEtAl1998}
{Ferland}, G.~J., {Korista}, K.~T., {Verner}, D.~A., {et~al.} 1998, \pasp, 110,
  761

\bibitem[{{Fogerty} {et~al.}(2016){Fogerty}, {Frank}, {Heitsch},
  {Carroll-Nellenback}, {Haig}, \& {Adams}}]{FogertyEtAl2016}
{Fogerty}, E., {Frank}, A., {Heitsch}, F., {et~al.} 2016, \mnras, 460, 2110

\bibitem[{{Fujimoto} {et~al.}(2014){Fujimoto}, {Tasker}, {Wakayama}, \&
  {Habe}}]{FujimotoEtAl2014}
{Fujimoto}, Y., {Tasker}, E.~J., {Wakayama}, M., \& {Habe}, A. 2014, \mnras,
  439, 936

\bibitem[{{Gazol} \& {Kim}(2013)}]{GazolKim2013}
{Gazol}, A., \& {Kim}, J. 2013, \apj, 765, 49

\bibitem[{{Ginsburg} {et~al.}(2013){Ginsburg}, {Federrath}, \&
  {Darling}}]{GinsburgFederrathDarling2013}
{Ginsburg}, A., {Federrath}, C., \& {Darling}, J. 2013, \apj, 779, 50

\bibitem[{{Glover} {et~al.}(2010){Glover}, {Federrath}, {Mac Low}, \&
  {Klessen}}]{GloverFederrathMacLowKlessen2010}
{Glover}, S.~C.~O., {Federrath}, C., {Mac Low}, M., \& {Klessen}, R.~S. 2010,
  \mnras, 404, 2

\bibitem[{{Goodman} {et~al.}(1998){Goodman}, {Barranco}, {Wilner}, \&
  {Heyer}}]{GoodmanEtAl1998}
{Goodman}, A.~A., {Barranco}, J.~A., {Wilner}, D.~J., \& {Heyer}, M.~H. 1998,
  \apj, 504, 223

\bibitem[{{Hennebelle} \& {Chabrier}(2011)}]{HennebelleChabrier2011}
{Hennebelle}, P., \& {Chabrier}, G. 2011, \apjl, 743, L29

\bibitem[{{Hennebelle} \& {Chabrier}(2013)}]{HennebelleChabrier2013}
---. 2013, \apj, 770, 150

\bibitem[{{Hennebelle} \& {Falgarone}(2012)}]{HennebelleFalgarone2012}
{Hennebelle}, P., \& {Falgarone}, E. 2012, \aapr, 20, 55

\bibitem[{{Hopkins}(2013)}]{Hopkins2013PDF}
{Hopkins}, P.~F. 2013, \mnras, 430, 1880

\bibitem[{{Hopkins} {et~al.}(2012){Hopkins}, {Quataert}, \&
  {Murray}}]{HopkinsQuataertMurray2012}
{Hopkins}, P.~F., {Quataert}, E., \& {Murray}, N. 2012, \mnras, 421, 3488

\bibitem[{{Howard} {et~al.}(2016){Howard}, {Pudritz}, \&
  {Harris}}]{HowardPudritzHarris2016}
{Howard}, C.~S., {Pudritz}, R.~E., \& {Harris}, W.~E. 2016, \mnras, 461, 2953

\bibitem[{{Hughes} {et~al.}(2010){Hughes}, {Wong}, {Ott}, {Muller}, {Pineda},
  {Mizuno}, {Bernard}, {Paradis}, {Maddison}, {Reach}, {Staveley-Smith},
  {Kawamura}, {Meixner}, {Kim}, {Onishi}, {Mizuno}, \&
  {Fukui}}]{HughesEtAl2010}
{Hughes}, A., {Wong}, T., {Ott}, J., {et~al.} 2010, \mnras, 406, 2065

\bibitem[{{Kainulainen} \& {Tan}(2013)}]{KainulainenTan2013}
{Kainulainen}, J., \& {Tan}, J.~C. 2013, \aap, 549, A53

\bibitem[{{Konstandin} {et~al.}(2012){Konstandin}, {Girichidis}, {Federrath},
  \& {Klessen}}]{KonstandinEtAl2012ApJ}
{Konstandin}, L., {Girichidis}, P., {Federrath}, C., \& {Klessen}, R.~S. 2012,
  \apj, 761, 149

\bibitem[{{K{\"o}rtgen} \& {Banerjee}(2015)}]{KoertgenBanerjee2015}
{K{\"o}rtgen}, B., \& {Banerjee}, R. 2015, \mnras, 451, 3340

\bibitem[{{K{\"o}rtgen} {et~al.}(2016){K{\"o}rtgen}, {Seifried}, {Banerjee},
  {V{\'a}zquez-Semadeni}, \& {Zamora-Avil{\'e}s}}]{KoertgenBanerjee2016}
{K{\"o}rtgen}, B., {Seifried}, D., {Banerjee}, R., {V{\'a}zquez-Semadeni}, E.,
  \& {Zamora-Avil{\'e}s}, M. 2016, \mnras, 459, 3460

\bibitem[{{Kritsuk} {et~al.}(2007){Kritsuk}, {Norman}, {Padoan}, \&
  {Wagner}}]{KritsukEtAl2007}
{Kritsuk}, A.~G., {Norman}, M.~L., {Padoan}, P., \& {Wagner}, R. 2007, \apj,
  665, 416

\bibitem[{{Krumholz} \& {McKee}(2005)}]{KrumholzMcKee2005}
{Krumholz}, M.~R., \& {McKee}, C.~F. 2005, \apj, 630, 250

\bibitem[{{Krumholz} \& {Tan}(2007)}]{KrumholzTan2007}
{Krumholz}, M.~R., \& {Tan}, J.~C. 2007, \apj, 654, 304

\bibitem[{{Larson}(1981)}]{Larson1981}
{Larson}, R.~B. 1981, \mnras, 194, 809

\bibitem[{{Mac Low}(1999)}]{MacLow1999}
{Mac Low}, M.-M. 1999, \apj, 524, 169

\bibitem[{{Mac Low} \& {Klessen}(2004)}]{MacLowKlessen2004}
{Mac Low}, M.-M., \& {Klessen}, R.~S. 2004, RvMP, 76, 125

\bibitem[{{Mac Low} {et~al.}(1998){Mac Low}, {Klessen}, {Burkert}, \&
  {Smith}}]{MacLowEtAl1998}
{Mac Low}, M.-M., {Klessen}, R.~S., {Burkert}, A., \& {Smith}, M.~D. 1998,
  PhRvL, 80, 2754

\bibitem[{{McKee} \& {Ostriker}(2007)}]{McKeeOstriker2007}
{McKee}, C.~F., \& {Ostriker}, E.~C. 2007, \araa, 45, 565

\bibitem[{{Molina} {et~al.}(2012){Molina}, {Glover}, {Federrath}, \&
  {Klessen}}]{MolinaEtAl2012}
{Molina}, F.~Z., {Glover}, S.~C.~O., {Federrath}, C., \& {Klessen}, R.~S. 2012,
  \mnras, 423, 2680

\bibitem[{{Murray}(2011)}]{Murray2011}
{Murray}, N. 2011, \apj, 729, 133

\bibitem[{{Nolan} {et~al.}(2015){Nolan}, {Federrath}, \&
  {Sutherland}}]{NolanFederrathSutherland2015}
{Nolan}, C.~A., {Federrath}, C., \& {Sutherland}, R.~S. 2015, \mnras, 451, 1380

\bibitem[{{Nordlund} \& {Padoan}(1999)}]{NordlundPadoan1999}
{Nordlund}, {\AA}., \& {Padoan}, P. 1999, in Interstellar Turbulence, ed.
  J.~{Franco} \& A.~{Carraminana}, 218

\bibitem[{{Ossenkopf} \& {Mac Low}(2002)}]{OssenkopfMacLow2002}
{Ossenkopf}, V., \& {Mac Low}, M.-M. 2002, \aap, 390, 307

\bibitem[{{Padoan}(1995)}]{Padoan1995}
{Padoan}, P. 1995, \mnras, 277, 377

\bibitem[{{Padoan} {et~al.}(2014){Padoan}, {Federrath}, {Chabrier}, {Evans},
  {Johnstone}, {J{\o}rgensen}, {McKee}, \& {Nordlund}}]{PadoanEtAl2014}
{Padoan}, P., {Federrath}, C., {Chabrier}, G., {et~al.} 2014, Protostars and
  Planets VI, 77

\bibitem[{{Padoan} \& {Nordlund}(2002)}]{PadoanNordlund2002}
{Padoan}, P., \& {Nordlund}, {\AA}. 2002, \apj, 576, 870

\bibitem[{{Padoan} \& {Nordlund}(2011)}]{PadoanNordlund2011}
---. 2011, \apj, 730, 40

\bibitem[{{Padoan} {et~al.}(1997){Padoan}, {Nordlund}, \&
  {Jones}}]{PadoanNordlundJones1997}
{Padoan}, P., {Nordlund}, {\AA}., \& {Jones}, B.~J.~T. 1997, \mnras, 288, 145

\bibitem[{{Padoan} {et~al.}(2016){Padoan}, {Pan}, {Haugb{\o}lle}, \&
  {Nordlund}}]{PadoanEtAl2016}
{Padoan}, P., {Pan}, L., {Haugb{\o}lle}, T., \& {Nordlund}, {\AA}. 2016, \apj,
  822, 11

\bibitem[{{Pan} {et~al.}(2016){Pan}, {Padoan}, {Haugb{\o}lle}, \&
  {Nordlund}}]{PanEtAl2016}
{Pan}, L., {Padoan}, P., {Haugb{\o}lle}, T., \& {Nordlund}, {\AA}. 2016, \apj,
  825, 30

\bibitem[{{Price} {et~al.}(2011){Price}, {Federrath}, \&
  {Brunt}}]{PriceFederrathBrunt2011}
{Price}, D.~J., {Federrath}, C., \& {Brunt}, C.~M. 2011, \apjl, 727, L21

\bibitem[{{Roman-Duval} {et~al.}(2010){Roman-Duval}, {Jackson}, {Heyer},
  {Rathborne}, \& {Simon}}]{RomanDuvalEtAl2010}
{Roman-Duval}, J., {Jackson}, J.~M., {Heyer}, M., {Rathborne}, J., \& {Simon},
  R. 2010, \apj, 723, 492

\bibitem[{{Salim} {et~al.}(2015){Salim}, {Federrath}, \&
  {Kewley}}]{SalimFederrathKewley2015}
{Salim}, D.~M., {Federrath}, C., \& {Kewley}, L.~J. 2015, \apjl, 806, L36

\bibitem[{{Stone} \& {Norman}(1992)}]{StoneNorman1992a}
{Stone}, J.~M., \& {Norman}, M.~L. 1992, \apjs, 80, 753

\bibitem[{{Stone} {et~al.}(1998){Stone}, {Ostriker}, \&
  {Gammie}}]{StoneOstrikerGammie1998}
{Stone}, J.~M., {Ostriker}, E.~C., \& {Gammie}, C.~F. 1998, \apjl, 508, L99

\bibitem[{{Sur} {et~al.}(2012){Sur}, {Federrath}, {Schleicher}, {Banerjee}, \&
  {Klessen}}]{SurEtAl2012}
{Sur}, S., {Federrath}, C., {Schleicher}, D.~R.~G., {Banerjee}, R., \&
  {Klessen}, R.~S. 2012, \mnras, 423, 3148

\bibitem[{{Tasker}(2011)}]{Tasker2011}
{Tasker}, E.~J. 2011, \apj, 730, 11

\bibitem[{{Tasker} \& {Tan}(2009)}]{TaskerTan2009}
{Tasker}, E.~J., \& {Tan}, J.~C. 2009, \apj, 700, 358

\bibitem[{{Tasker} {et~al.}(2015){Tasker}, {Wadsley}, \&
  {Pudritz}}]{TaskerWadsleyPudritz2015}
{Tasker}, E.~J., {Wadsley}, J., \& {Pudritz}, R. 2015, \apj, 801, 33

\bibitem[{{Truelove} {et~al.}(1997){Truelove}, {Klein}, {McKee}, {Holliman},
  {Howell}, \& {Greenough}}]{TrueloveEtAl1997}
{Truelove}, J.~K., {Klein}, R.~I., {McKee}, C.~F., {et~al.} 1997, \apjl, 489,
  L179

\bibitem[{{Turk} {et~al.}(2011){Turk}, {Smith}, {Oishi}, {Skory}, {Skillman},
  {Abel}, \& {Norman}}]{TurkEtAl2011}
{Turk}, M.~J., {Smith}, B.~D., {Oishi}, J.~S., {et~al.} 2011, \apjs, 192, 9

\bibitem[{{V{\'a}zquez-Semadeni}(1994)}]{Vazquez1994}
{V{\'a}zquez-Semadeni}, E. 1994, \apj, 423, 681

\bibitem[{{V{\'a}zquez-Semadeni} {et~al.}(2003){V{\'a}zquez-Semadeni},
  {Ballesteros-Paredes}, \& {Klessen}}]{VazquezBallesterosKlessen2003}
{V{\'a}zquez-Semadeni}, E., {Ballesteros-Paredes}, J., \& {Klessen}, R.~S.
  2003, \apjl, 585, L131

\bibitem[{{Williams} \& {McKee}(1997)}]{WilliamsMcKee1997}
{Williams}, J.~P., \& {McKee}, C.~F. 1997, \apj, 476, 166

\bibitem[{{Zuckerman} \& {Evans}(1974)}]{ZuckermanEvans1974}
{Zuckerman}, B., \& {Evans}, II, N.~J. 1974, \apjl, 192, L149

\end{thebibliography}

\end{document}